# Melting of hexane monolayers adsorbed on graphite: the role of domains and defect formation.

C. Wexler[1], L. Firlej[1,2], B. Kuchta[1,3] and M.W. Roth[4]

[2]*University of Missouri, Department of Physics and Astronomy, Columbia, MO 65211*
[2]*LCVN, Université Montpellier 2, 34095 Montpellier, France*
[3]*Laboratoire Chimie Provence, Université de Provence, 13396 Marseille, France*
[4]*University of Northern Iowa, Department of Physics, Cedar Falls, IA 50614*

(March 4, 2009)

**Abstract**

We present the first large-scale molecular dynamics simulations of hexane on graphite that completely reproduces all experimental features of the melting transition. The canonical ensemble simulations required and used the most realistic model of the system: (i) fully atomistic representation of hexane; (ii) explicit site-by-site interaction with carbon atoms in graphite; (iii) CHARMM force field with carefully chosen adjustable parameters of non-bonded interaction; (iv) numerous $\geq 100$ ns runs, requiring a total computation time of ca. 10 CPU-years. This has allowed us to determine correctly the mechanism of the transition: molecular reorientation within lamellae without perturbation of the overall adsorbed film structure. We observe that the melted phase has a dynamically reorienting domain-type structure whose orientations reflect that of graphite.

## I. Introduction

Hexane is the shortest alkane whose flexibility has any significant impact on its dynamics. Its behavior on a graphite substrate has been extensively studied both experimentally [1-5] and computationally [4-13]. Neutron scattering and X-ray diffraction reveal that, at near monolayer coverage, the system transits from a herringbone solid into a rectangular solid/liquid coexistence region as the temperature is raised, finally melting at temperatures ~ 170 K [1-5].

Even with a considerable body of computational work [4-13], there remain poorly understood elements of this interesting system. In particular, the detailed mechanism of the melting transition and the effect that molecular stiffness has on it has not been elucidated. There are at least three reasons for that. First, the issue of simulating a formally complete monolayer at zero spreading pressure has never been investigated. Although believed to be modest in previous simulations [6-8,10,12] any planar stress present in phase transition simulations can, and in fact does, dramatically affect the system dynamics [11]. Second, molecular flexibility was not properly accounted for in previous studies. Third, when the flexibility is properly modeled, it creates the need for surprisingly long simulation times (equilibration plus production runs), on the order of 50 to 220 ns that have never been carried out. The impetus of the work reported here are the issues presented above, and it entails extensive massively parallel computer simulations which, due in large part to the long equilibration times required for the system and the need for robust statistics, have taken ca. 10 CPU-years.

## II. Computational aspects

The all-atom description of hexane molecule used in the present study comes from the Brookhaven Protein Data Bank (PDB) [16]. The initial low-temperature configuration has an important departure from the previous ones [6-12]. There are $N = 104$ hexane molecules in a herringbone arrangement atop a six-layer 68.16 Å x 68.88 Å graphite structure. Such a structure has virtually no spreading pressure. Molecular Dynamic simulations were run for a total of 40 ns of stabilization followed by at least 100 ns of production runs. All other simulation parameters, including the standard CHARMM22 interaction parameters [14], were exactly the same as in a previous paper [12]. Here we focus only on one aspect of the model's molecular flexibility: how to correctly account for the intramolecular nonbonded van der Waals energies and Coulomb electrostatic potential. Within CHARMM22 force field, these interactions for 1-3 pairs (first through third neighbors on the same molecule) are not included, and those for 1-5 and beyond are fully included. However, one can scale the 1-4 interactions, and drastically modify the molecular stiffness. Many simulations employ a generic value of the scaling factor $SF = 0.5$, however, the assignment of such a value is poorly understood. In a recent study we have shown [13] that the optimal value of scaling factor for alkanes varies as a function of the chain lengths and we have determined the optimal SF value for hexane to be $SF \cong 0.8$. This value has been used in the present simulations.

## III. Results and Discussion

The most important result emerging from the simulation is the mechanism of melting which is consistent with the experimental data [9] and was not reproduced in the previous papers [10,12].

Figure 1 shows the order parameters, as used before [12]. There is a profound and synchronous loss of order in the system at $T = 170$ K, which is attributed to melting. The pair correlation function (not shown) also confirms that the system undergoes a sharp melting transition. It is noteworthy that the behavior of all the order parameters in Figure 1 illustrate that the system melts directly from a herringbone to the fluid, in contrast to all previous computer simulations where a nematic mesophase is observed [7,8,10-12].

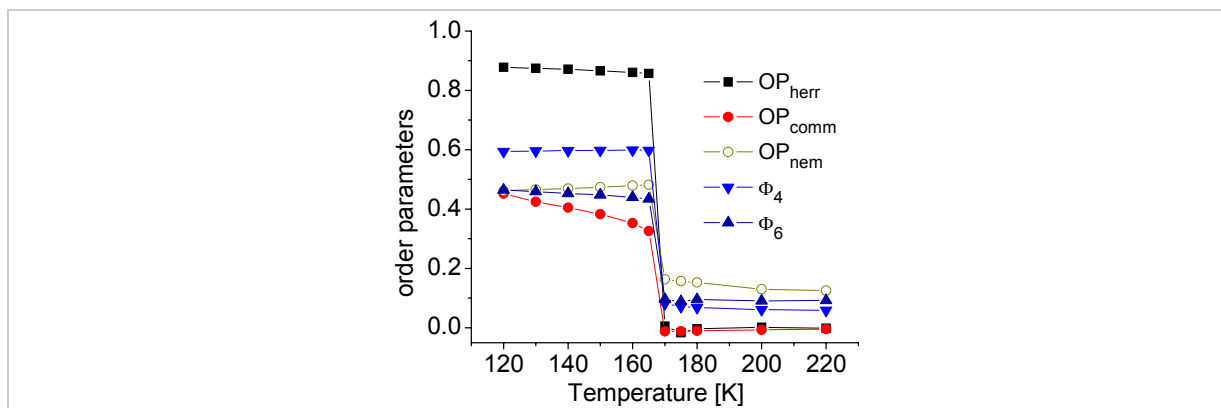

**Figure 1**. Evolution of system structure (order parameters) with temperature: $OP_{herr}$, herringbone structure; $OP_{comm}$, commensurability of the adlayer with the substrate; $OP_{nem}$, nematic-like phase; $\Phi_4$, square 2D structure; and $\Phi_6$, triangular 2D structure. Further details are given in Ref. 12.

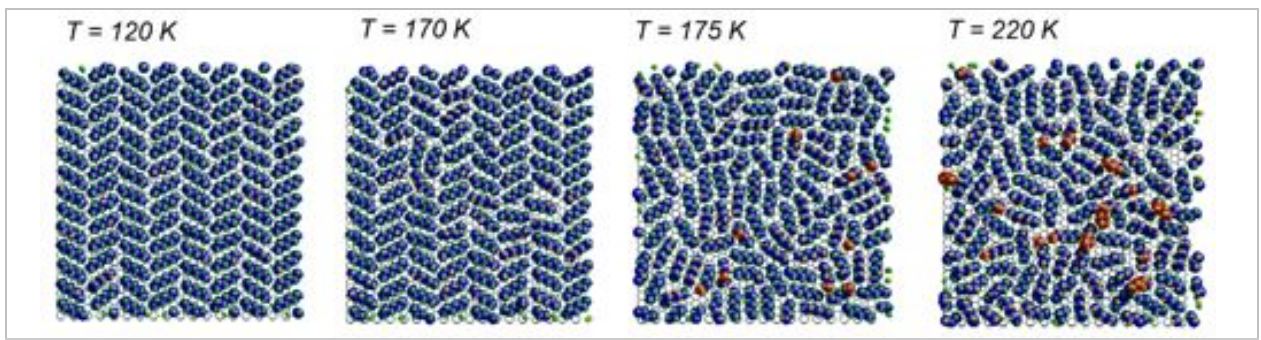

**Figure 2.** Configuration snapshots from $T = 120K$ to 220K. Hexane carbons are blue and hydrogens are green for heights under 5 Å and become green and yellow, respectively if they are higher.

Figure 2 shows typical configurations of the layer at various temperatures. The formation of relatively large domains at the melting is evident; the domains size decreases with increasing temperature. The domains start to form before melting as the defects within the lamella ($OP_{nem}$ has a small positive slope which increases up until the loss of order at melting) and persist in the fluid phase ($OP_{nem}$ does not vanish but stabilizes at the value ~0.2). Moreover, the azimuthal angle distributions (not shown) are consistent with this picture: even in the liquid a residual orientational order remains, suggesting that the dynamically forming domains orient themselves so as to reflect the symmetry of the underlying substrate.

The tilting of, and rotation about the long molecular axis also exhibit distinct signatures of melting. Figure 3 shows distributions of microscopic rolling angle $\Phi$ (the angle between the plane formed by each three-body carbon segment in the molecule and the graphite substrate) [10,12]. In the solid there is virtually no population of out-of-plane rolling angles, which is in excellent agreement with experiment [1-3]. A peak grows at $\Phi = 90º$ as the temperature increases; it indicates that prior to melting molecules start to roll on their sides to create in-plane room and initiate melting. Finally, due to the long (100 ns) runs involved, fine points of the roll angle distributions can be detected, such as a lifting at 30º and 150º during melting. We interpret these changes in the distributions to indicate the presence of gauche defects coupled with rolling.

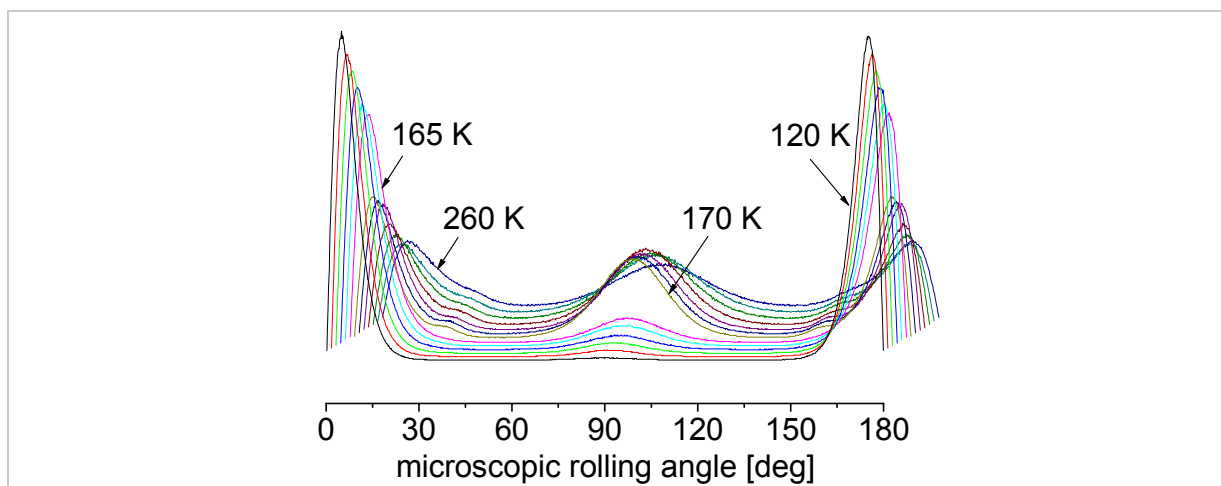

**Figure 3.** Microscopic roll angle probability distributions $P(\Phi)$ at various temperatures.

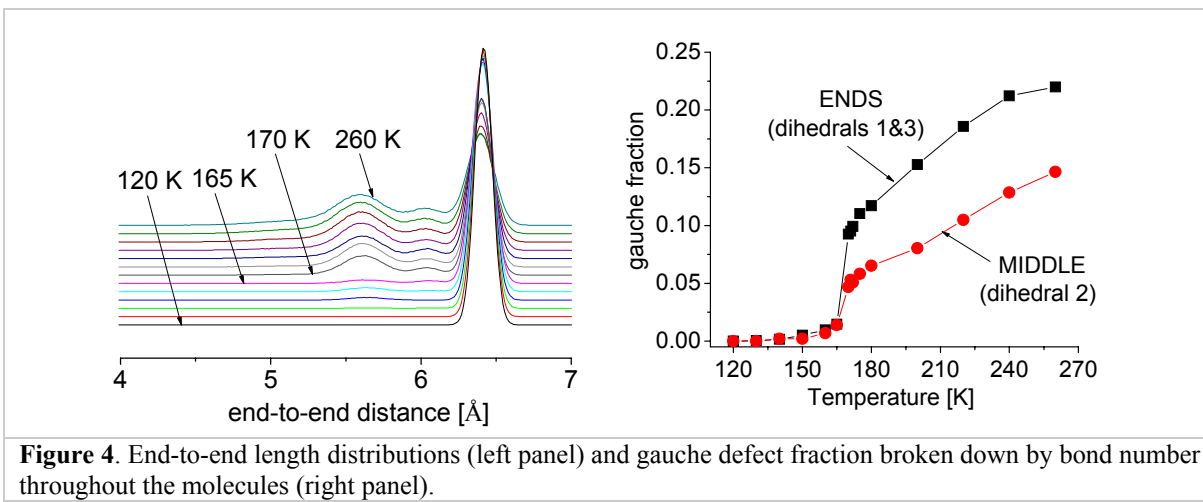

**Figure 4**. End-to-end length distributions (left panel) and gauche defect fraction broken down by bond number throughout the molecules (right panel).

Figure 4 shows end-to-end length distributions and gauche defect fraction. The end-to-end distance distributions at various temperatures confirm the presence of molecular distortion with the emergence of the broad peak centered a little higher than 5.5 Å. This proves that melting is strongly determined by the possibility of molecular deformations. This unique feature of hexane can be seen by the breakdown of gauche defects by bond number in Fig. 4. Gauche defects do not propagate from the ends of the molecules inward: they occur over the entire molecule at once. This, in combination with in-plane space creation by rolling, drives the melting transition. This makes it clear, for the first time, why the melting transition of hexane adsorbed on graphite is so sharp compared to other systems.

**Conclusions**

There are two main conclusions that complete our previous studies of hexane layers adsorbed on graphite.

First, when the structure of the system is correctly modeled (the optimal scaling factor SF = 0.8 is used [13]), the monolayer melts directly from the herringbone solid to a fluid with residual local order reflecting the symmetry of the underlying substrate. The melting dynamics proceeds by domain formation resulting from gauche deformation of molecules in concert with rolling and tilting. It is the first time that practically all experimental characteristics of the hexane melting have been reproduced and the phase transition mechanism explained with a help of numerical simulations. Additionally, we show that the structure of melted phase consists in a network of rectangular centered islands mobile in the fluid, in accordance with the scattering data [1-5,9].

Second, as the molecules are flexible, equilibration times for the system can be surprisingly long and one can get a false sense of the simulation having stabilized if the statistics of the results is not properly analyzed. A significant computational effort is thus required, even for this relatively simple model system.

It was not surprising that the interaction model played the decisive role in getting the correct melting temperature. However, it is not intuitively evident that the scaling of the 1-4 nonbonded interactions affects dramatically both the melting temperature and the mechanism of melting. This feature, emphasized in this paper and also observed in other alkanes [13,17], is likely important in simulations of any flexible molecule. This hypothesis needs more studies to be verified.

**Acknowledgements**

The authors acknowledge useful discussions with H. Taub. This material is based upon work supported in part by the Department of Energy under Award Number DE-FG02-07ER46411. Acknowledgment is made to the Donors of The American Chemical Society Petroleum Research Fund (PRF43277-B5). Computational support was provided by the University of Missouri Bioinformatics Consortium.